\pdfoutput=1
\documentclass[11pt]{article}

\usepackage[preprint]{acl}
\usepackage{times}
\usepackage{url}
\usepackage{latexsym}
\usepackage{cuted}
\usepackage{enumitem}
\usepackage{tabularx}
\usepackage[T1]{fontenc}

\usepackage[most]{tcolorbox}
\usepackage{xcolor}
\definecolor{mygreen}{RGB}{146,209,79}
\newtcbox{\gbox}{on line,
colback=mygreen,
  colframe=black,        % black border
  arc=2mm,               % roundness of corners
  boxrule=0.4pt,         % border thickness
  left=0.3mm,right=0.3mm,top=0.1mm,bottom=0.1mm % padding
}
\tcbset{
  colback=white,
  colframe=black,
  boxrule=0.6pt,
  arc=2pt,
  left=4pt,right=4pt,top=3pt,bottom=3pt,
  boxsep=2pt,
}

\newcommand{\goodtok}[1]{\colorbox{blue!15}{\textcolor{blue!75!black}{\textbf{#1}}}}
\newcommand{\badtok}[1]{\colorbox{pink!25}{\textcolor{red!75!black}{\textbf{#1}}}}

\lstdefinestyle{reqallocBase}{
  basicstyle=\ttfamily\fontsize{10.5pt}{10.5pt}\selectfont,
  columns=fullflexible,
  keepspaces=true,
  showstringspaces=false,
  breaklines=true
}

\lstdefinestyle{reqallocCorrect}{
  style=reqallocBase,
  literate=
    {::}{{{\goodtok{::}}}}2
}

\lstdefinestyle{reqallocWrong}{
  style=reqallocBase,
  literate=
    {.}{{{\badtok{.}}}}1
}

\usepackage[utf8]{inputenc}

\usepackage{microtype}

\usepackage{inconsolata}

\usepackage{graphicx}
\usepackage{microtype}
\usepackage{subcaption} 
\usepackage{booktabs} % for professional tables
\usepackage{comment} 

\newcommand{\CommentedText}[1]{}

\usepackage{amsmath}
\usepackage{amssymb}
\usepackage{mathtools}
\usepackage{amsthm}
\usepackage{algorithm}
\usepackage{algorithmic}

\theoremstyle{plain}

\theoremstyle{definition}

\theoremstyle{remark}

\title{Workflow-Level Design Principles for Trustworthy GenAI \\in Automotive System Engineering}

\author{
  \textbf{Chih-Hong Cheng\textsuperscript{1}},
  \textbf{Brian Hsuan-Cheng Liao\textsuperscript{2}}, 
  \textbf{Adam Molin\textsuperscript{2}},
  \textbf{Hasan Esen\textsuperscript{2}}
  \\
    \textsuperscript{1}Carl von Ossietzky University of Oldenburg, Oldenburg, Germany\\
    \textsuperscript{2}DENSO AUTOMOTIVE Deutschland Germany, Eching, Germany
}

\begin{document}
\maketitle
\begin{abstract}

The adoption of large language models in safety-critical system engineering is constrained by trustworthiness, traceability, and alignment with established verification practices. We propose workflow-level design principles for trustworthy GenAI integration and demonstrate them in an end-to-end automotive pipeline, from requirement delta identification to SysML v2 architecture update and re-testing. First, we show that monolithic (``big-bang'') prompting misses critical changes in large specifications, while section-wise decomposition with diversity sampling and lightweight NLP sanity checks improves completeness and correctness. Then, we propagate requirement deltas into SysML~v2 models and validate updates via compilation and static analysis. Additionally, we ensure traceable regression testing by generating test cases through explicit mappings from specification variables to architectural ports and states, providing practical safeguards for GenAI used in safety-critical automotive engineering.

\end{abstract}

\section{Introduction}

Automotive systems are rapidly increasing in complexity, spanning intelligent thermal management, electrified powertrains, and advanced driver assistance functions. This growth exacerbates two persistent bottlenecks: (i) requirements (customer specifications, regulations, and process standards) evolve continuously, forcing repeated delta identification and interpretation for downstream design and validation; and (ii) as systems scale, end-to-end traceability from requirements to architecture, implementation, and verification becomes harder, slowing iteration and increasing the risk of inconsistent updates.

Generative AI offers strong leverage when grounded in external evidence and iterative reasoning. Retrieval-Augmented Generation (RAG) mitigates hallucinations by conditioning outputs on retrieved passages~\cite{lewis2020rag}, while recent graph-based variants better capture complex structural dependencies~\cite{edge2024global,he2024gretriever,gutierrez2024hipporag}. Agentic frameworks such as ReAct and Tree of Thoughts extend these capabilities by interleaving reasoning with tool execution and multi-path planning~\cite{yao2023react,yao2024tree}. These architectures have demonstrated potential in automated software engineering and testing loops~\cite{yang2024sweagent,wang2025automating}. To improve reliability, techniques like Reflexion and self-consistency mitigate brittleness through iterative feedback and consensus sampling~\cite{shinn2023reflexion,wang2022selfconsistency}, while formal verification provides machine-checkable guarantees~\cite{bhatia2024verified,sevenhuijsen25generating}. 
However, in automotive system engineering, these techniques are often assembled ad hoc, and guidance for structuring GenAI workflows that remain traceable and verification-compatible is limited.

We address this gap by contributing workflow-level design principles that favor modular, checkable pipelines over monolithic prompting: (A) define fine-grained tasks with explicit inputs/outputs; (B) implement suitable steps with classical or symbolic methods to improve repeatability and testability; (C) strengthen neural steps via safety engineering principles such as diversification and consensus; and (D) validate neural outputs with necessary-condition checkers. Importantly, parts of the resulting toolchain are already deployed and used in an industrial setting, providing early evidence of substantial efficiency gains in turnaround time for requirements-change review. Our approach is orthogonal to simply using bigger models: we focus on workflow structure (modularization, checks, and tool-backed validation) that improves reliability and traceability regardless of model scale.

We apply these principles to two stages of a single end-to-end engineering pipeline, aligned with the two introduced bottlenecks: (i) requirements version comparison and delta identification, featuring section-wise processing with persisted intermediate artifacts, diversified retrieval, and sanity checks; and (ii) model-based architecture update and validation, featuring incremental SysML~v2 deltas with compilation/static analysis. In particular, we also regenerate verification artifacts traceably by constraining test-case generation to explicit mappings between specification variables and architectural ports/states, demonstrating practical safeguards for deploying GenAI in safety-critical automotive system engineering.

\section{Vision}

\begin{figure}[t]
  \centering
  \includegraphics[width=\columnwidth]{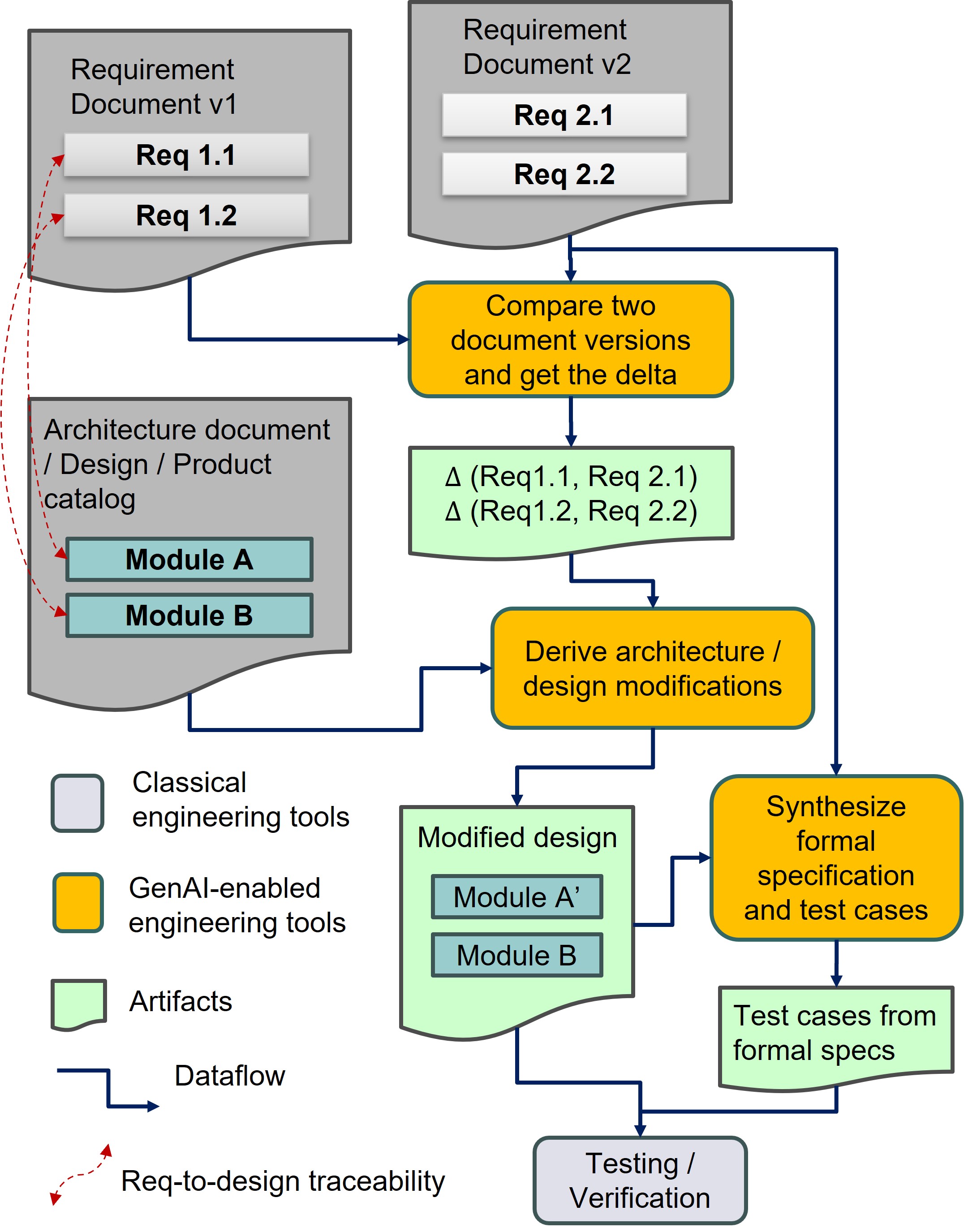}
 \caption{GenAI-assisted workflow from requirement updates to architecture changes and verification.}
  \label{fig:vision}
\end{figure}

Fig.~\ref{fig:vision} depicts the workflow we envision for a systems engineering organization when receiving updated customer requirements (e.g., when a Tier~1 automotive supplier receives an updated RfQ or regulatory document). The new requirements document (v2) is compared against the previous baseline (v1) to produce a \emph{structured delta} (additions, removals, and modifications with traceable evidence). This delta is then interpreted against the current design baseline; it may range from a company component catalog to a platform/reference implementation and, ultimately, a project-specific architecture model, in order to derive concrete, incremental design modifications (e.g., Module A $\rightarrow$ Module A$'$). This reflects industrial reality, where redesigning from scratch is rarely feasible, and the reuse of existing designs is economically essential.

The updated design is re-validated by generating verification artifacts (e.g., formal specifications and test cases or test plans) that are explicitly grounded in the architecture via mappings between specification variables and model ports/states/parameters, enabling systematic regression testing. GenAI can accelerate delta extraction, impact analysis, and test synthesis, but the final validation may still require non-software evidence, such as hardware-in-the-loop and vehicle-level tests for cyber-physical requirements.

\section{Towards Trustworthy GenAI for System Engineering}

\subsection{Issues with GenAI}

We started our journey in implementing the ``analysis of the requirement delta'' as depicted in Fig.~\ref{fig:vision}. 
While an LLM can perform analysis smoothly for two simple requirements in markdown or text files with only a few requirements, our use case involves complex documents with multiple sections that can easily reach hundreds of pages. The business demand also requires the tool to avoid overlooking any critical aspects. 
We have experimented with multiple variations: The initial monolithic workflow attempted to do the following three things all within a single agentic LLM call with additional post-processing to coerce JSON outputs (including an ``Undetermined'' fallback).
\begin{itemize}[nosep]%[label=(\Alph*), nosep]
    \item Summarize a section from the newer document version (v2).
    \item  Retrieve relevant sections from the older version (v1) based on the section title and section text of v2. 
    \item  List specification changes.
\end{itemize}

In our evaluation, this ``big-bang'' or ``do-everything'' setup proved fragile: it overloaded a single neural component with multiple responsibilities (reasoning, memory, structured output, and repeated criterion checks), creating hallucinations, omitting instructions, making prompt tuning brittle and prone to side effects. Such a situation is generic, as it also occurs in design modification or test case generation. \emph{Note that although changing to larger models helps, the problems are never eliminated}. This leads to a rethinking of the overall approach that is detailed in subsequent sections. 

\vspace{2mm}
\noindent\textbf{(Example)} Because company-internal use cases are highly confidential, we use a publicly available proxy task to illustrate the approach: an LLM-assisted comparison of Automotive SPICE (ASPICE) versions 3.1 and 4.0.\footnote{\url{https://en.wikipedia.org/wiki/Automotive_SPICE}}
 ASPICE provides a process reference model and process assessment model that function as de facto standards in the automotive domain. Since each version exceeds 100 pages, naive monolithic prompting tends to miss many changes, motivating our modular and checker-based workflow.

\subsection{Design Principles for Trustworthy GenAI}

Our key lessons learned are that, to make GenAI-enabled system engineering more trustworthy, \emph{in addition to using larger models}, it is necessary to move away from a monolithic prompting approach (including tool calling capabilities) toward a modular, checkable workflow. We have summarized our approach as follows.
\begin{enumerate}[label=(\Alph*), nosep]
    \item Design a fine-grained structural workflow  characterizing all activities. For each activity, define the corresponding inputs and outputs.
    \item Consider what parts in the workflow can use symbolic methods / classical algorithms without introducing neural components. Precisely, for modules where the classical algorithm shines, do not wrap it inside an LLM-enabled agent to ensure correctness, repeatability, and testability. 
    \item Increase the output quality of neural components via classical architectural tactics in safety engineering, such as diversity, consensus voting, and/or checkpoint rollbacks.
    \item Examine the quality of the output from neural components via checking against \emph{necessary conditions for correctness}. This enables the introduction of classical algorithms that can serve as \emph{filters} on potentially incorrect results. 
\end{enumerate}

\vspace{1mm}
In the remaining part of the paper, we illustrate how the principles are implemented in our system.

\section{Rigorous Workflow for Requirement Document Version Comparison}

\begin{figure}[t]
  \centering
  \includegraphics[width=\columnwidth]{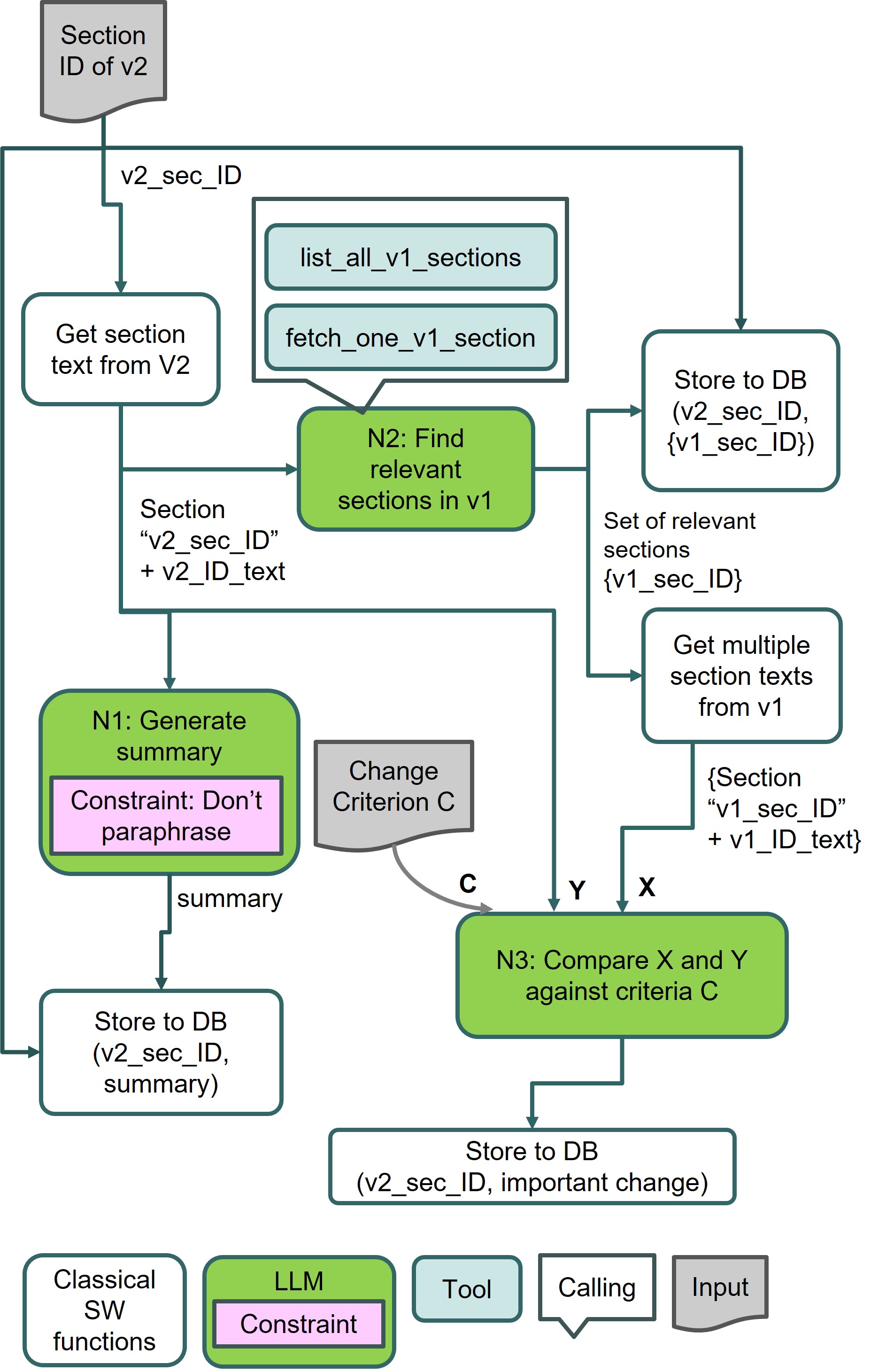}
 \caption{A revised workflow for comparing complicated documents while increasing trustworthiness.}
  \label{fig:req.workflow}
\end{figure}

Instead of treating version comparison as a monolithic ``document-to-document'' diff, we perform a \emph{section-by-section} analysis. Concretely, we iterate over sections in the newer version (v2) and, for each v2 section, determine what has changed compared to the most relevant content in the previous version. This design matches how requirements are authored and reviewed in practice (engineers navigate by section IDs/headings), and it provides a stable unit for caching, auditing, and verification.

\vspace{-2mm}
\subsection{PDF Sectionization and Canonical Storage (A)}
\vspace{-1mm}

To enable section-level comparison, we first normalize each input PDF (v1 and v2) into a shared \emph{section representation}. We create a sectionization component, implemented either with classical heuristics (e.g., heading numbering patterns, typography/layout cues as in PdfPlumber\footnote{\url{https://github.com/jsvine/pdfplumber}}) or neural document parsing (e.g., MinerU\footnote{\url{https://github.com/opendatalab/MinerU}}), to extract: (i) a stable \texttt{section\_id} (derived from numbering/title and its hierarchy), (ii) cleaned section text, and (iii) optional metadata (e.g., heading string, parent section). We then store all extracted sections from both versions in a canonical database schema. This decouples expensive PDF parsing from downstream comparison, and allows repeated evaluation of prompts/retrieval/criteria without reprocessing PDFs. Note that, as is commonly seen in RfQs, the section ID and titles can differ between versions. 

\vspace{-1mm}
\subsection{Decomposition into Neural and Classical Micro-Tasks (A,B)}
\vspace{-1mm}

Given the sectionized corpus, the workflow in Fig.~\ref{fig:req.workflow} is applied independently to each v2 section:
\begin{itemize}[leftmargin=1em, nosep]
  \item \gbox{N1} \textbf{Generate summary} for a v2 section under an explicit constraint. We add a constraint requiring the LLM \textbf{not} to paraphrase (reasons discussed later), which also facilitates evidence-based checks.
  \item \gbox{N2} \textbf{Find relevant sections in v1} by calling tools over the stored v1 sections (e.g., listing all section IDs via \texttt{list\_all\_v1\_sections} and fetching content via \texttt{fetch\_one\_v1\_section}). The output is a small set of candidate v1 section IDs aligned to the current v2 section.
  \item \gbox{N3} \textbf{Compare v2 section text (X) against retrieved v1 section text(s) (Y)} under a selected \emph{change criterion} $C$, and output structured changes with supporting evidence excerpts (pieces of sentences in the document).
\end{itemize}

The information flow across the neural components is in many parts handled or supported by classical algorithms. For example, an alternative \gbox{N2} design could directly have the LLM (i.e., the neural component) output the contents of a v1 section that it fetches and presumes relevant.  However, it suffices to have the LLM produce the set of section IDs in v1 (i.e., \texttt{\{v1\_sec\_ID\}} in Fig.~\ref{fig:req.workflow}), and we can use a classical algorithm to prepare the sections based on the set of relevant v1 section IDs.

Furthermore, we persist intermediate artifacts (i.e., generated summaries, retrieved v1 section IDs, identified criterion-wise changes) to support isolated debugging, prompt reuse, and targeted improvements to individual modules. Similarly, all these persistences can be done without being wrapped by LLMs.

\vspace{-1mm}
\subsection{Robustifying Neural Outputs via Diversification and Unification (C)}

To mitigate omissions  in retrieving relevant v1 sections, we introduced multiple variants of the section-retrieval step (\gbox{N2}) in Fig.~\ref{fig:req.workflow}:
\begin{itemize}[leftmargin=1em, nosep]
  \item \texttt{find\_relevant\_sec()}: Use a deterministic single-model call.
  \item \texttt{find\_relevant\_sec\_redundant()}: Apply the same model with different random seeds, followed by merging results. 
  \item \texttt{find\_relevant\_sec\_different\_LLMs()}: Apply different models then merge results. 
\end{itemize}

\noindent Our design decision to have N2 output only section IDs in v1 makes creating the consensus decision fundamentally simpler. One can either take the \emph{majority vote}, or act safely by taking the \emph{union of the proposed section IDs}. Creating such a resolution can be done with a simple classical algorithm without involving a neural model such as LLM. 

\begin{figure}
    \centering
    \includegraphics[width=\linewidth]{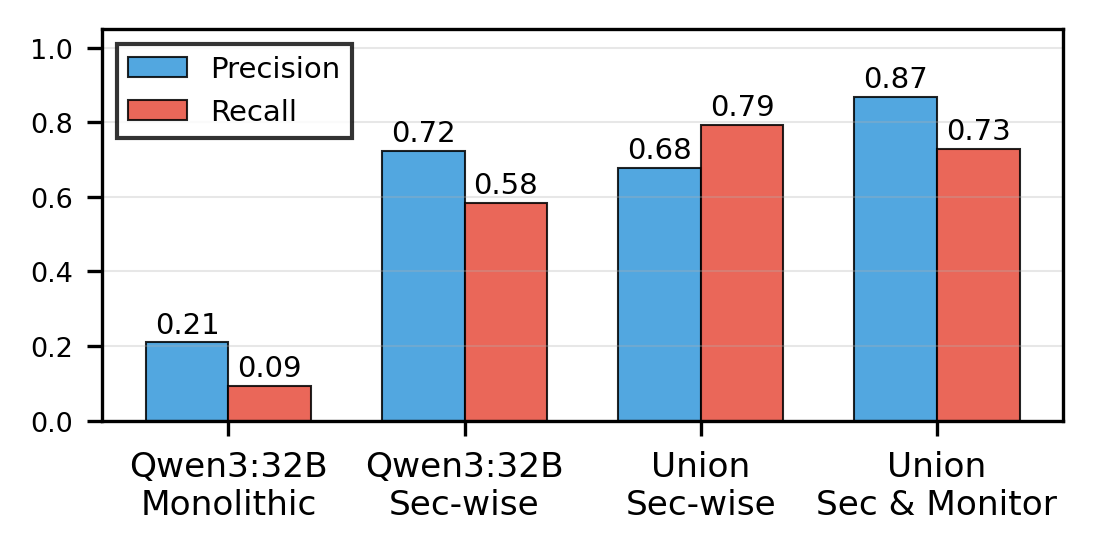}
    \vspace{-8mm}
    \caption{Precision and recall per model setting in retrieving relevant sections across ASPICE v3.1 and v4. \texttt{Union} refers to combining the predictions of Qwen3:32B, Nemotron3:30B, and GPT-OSS:20B.}
    \label{fig:benchmark}
      % \vspace{-2mm}
\end{figure}

\vspace{1mm}
\noindent\textbf{(Example)} We run experiments with \texttt{Qwen3:32B} to retrieve relevant sections across ASPICE v3.1 and v4, over the whole document. The predictions are evaluated against an oracle revised by engineers. As Fig.~\ref{fig:benchmark} shows, section-wise processing significantly outperforms the monolithic baseline. Unifying \texttt{Qwen3:32B} with two additional open-source models (\texttt{NVIDIA-Nemotron-3-Nano:30b}, \texttt{gpt-oss:20b}) around a similar parameter size further raises recall to 0.79.

\vspace{1mm}

Finally, for comparing texts from sections and extracting differences (\gbox{N3} in Fig.~\ref{fig:req.workflow}), to reduce ambiguity around what constitutes a ``change,'' in addition to simply asking for enumerating changes, one can also repeatedly call the neural component with explicitly stated change criterion mentioned by domain experts. Extracted changes subject to each criterion can then be merged separately again without the use of neural components. 

\vspace{1mm}
\noindent\textbf{(Example)} For the example of APICE version comparison, it includes cases such as scope changes (add/remove entities), modality changes (``shall'' vs.\ ``may''), numeric/tolerance changes, newly added or removed mandatory requirements, newly introduced definitions, renamed tools/technologies, and terminology changes that introduce ambiguity. Thanks to having only relevant sections being offered, the context for LLM is fundamentally shorter, making the result more precise in comparison to feeding into the complete two documents (each has around $150$ pages).

\vspace{-1mm}
\subsection{Designing Checkers (D)}
\vspace{-1mm}

With well-defined I/O for each module, we defined ``contracts'' (necessary conditions) that can be monitored using classical methods. Here we list some of the checks being implemented.

\noindent\textbf{Relevance sanity check for \gbox{N2}:} Use classical similarity measures such as TF--IDF~\cite{salton1975vector,salton1988term}. Given two text segments $A$ and $B$, denote their TF--IDF cosine similarity to be $\mathrm{s}(A,B)$. Given text $T^{v2}_{ID_i}$ from v2 with section $ID_i$, We envision that for the \gbox{N2} identified relevant v1 sections ${ID_j}$, their computed cosine similarity should be larger than other randomly selected sections in v1.  
  In other words, for $K$ randomly selected $ID_{r_1}, \ldots, ID_{r_K}$ section IDs in v1 that do not fall inside \gbox{N2} considered v1 sections, they should satisfy Eq.~\eqref{eq:td}. Our sanity checker then enables a warning when $\alpha\%$ or more of the inequality constraints are violated. 
  
%\vspace{-2mm}

{\setlength{\abovedisplayskip}{4pt}
 \setlength{\belowdisplayskip}{6pt}
 \setlength{\abovedisplayshortskip}{2pt}
 \setlength{\belowdisplayshortskip}{2pt}
\begin{equation}
\begin{aligned}
\mathrm{s}(T^{v2}_{ID_i},T^{v1}_{ID_j}) > \mathrm{s}(T^{v2}_{ID_i},T^{v1}_{ID_{r_1}}) \\
          \cdots \\
\mathrm{s}(T^{v2}_{ID_i},T^{v1}_{ID_j}) > \mathrm{s}(T^{v2}_{ID_i},T^{v1}_{ID_{r_K}})
\end{aligned}
\label{eq:td}
\end{equation}
}

\vspace{1mm}
\noindent\textbf{(Example)} Building on Fig.~\ref{fig:benchmark}, we set $\alpha=30$ to prune low-quality union predictions, increasing precision to 0.87 with only a small recall drop.
\vspace{1mm}

\noindent\textbf{Summary check for \gbox{N1}:} Similarly, as in the design of \gbox{N1}, we explicitly demands the neural component not to paraphrase, classical methods such as TF--IDF cosine similarity can also be used; it raises a warning if the computed similarity is lower than a certain threshold~$\beta$. 

\noindent\textbf{Validity checking for \gbox{N3}} In our workflow, we demand \gbox{N3} to produce a list of changes using three tuples $\langle ccd, ev_{v1}, ev_{v2} \rangle$.
  \begin{itemize}[leftmargin=1.2em, nosep]
      \item $ccd$ is the concrete change description between v1 and v2. 
      \item $ev_{v1}$ is the verbatim excerpt in v1 section text, or ``\texttt{not in v1}'' when the change implies a new concept. 
      \item $ev_{v2}$ is the verbatim excerpt in v2 section text, or ``\texttt{not in v2}'' when the change implies a concept deletion. 
  \end{itemize}

Such a tuple structure enables multiple checks after the LLM produces the output.

\begin{enumerate}[label=(\roman*), leftmargin=1.2em, nosep]
    \item If $ev_{v1}$ equals ``\texttt{not in v1}'', this implies that $ev_{v2}$ is not in the section text of v1 ($T^{v1}_{ID_j}$). Therefore, $\mathrm{s}(T^{v1}_{ID_j}, ev_{v2})$ should return a small value. Based on this, our checker issues a warning when $\mathrm{s}(T^{v1}_{ID_j}, ev_{v2})$ exceeds a threshold~$\kappa$ (indicating concept overlap). 
    
    \item Dually, if $ev_{v2}$ equals ``\texttt{not in v2}'', our checker raises a warning when $\mathrm{s}(T^{v2}_{ID_i}, ev_{v1})$ is larger than a certain threshold~$\kappa$. 

    \item We also apply Natural Language Inferencing (NLI) techniques~\cite{maccartney2009natural,bowman2015large,devlin2019bert,zhuang2021robustly} to examine whether premises $ev_{v1}$ and $ev_{v2}$ jointly entail the hypothesis~$ccd$. When NLI reports ``\texttt{contradiction}'', a warning is raised. 
\end{enumerate}

It is crucial to understand that these checks do not eliminate errors completely, but they provide first-principled monitors that flag likely failures and enable engineers to manually intervene.

\vspace{-2mm}
\section{Design Modification and Test Case Regeneration}
\vspace{-1mm}

Once the requirement changes are identified, the next step is to determine how the design should be updated. In the automotive domain, both the architecture and the requirement-to-design traceability are typically made explicit, and engineers generally prefer \emph{mild} (incremental) modifications over redesigning from scratch. We illustrate this stage using a simplified Automatic Emergency Braking (AEB) example. In particular, we maintain the requirements, the corresponding design model, and the requirement-to-design traceability in SysML v2, which is the next-generation modeling language under active development\footnote{\url{https://www.omg.org/sysml/sysmlv2/}}. 

\vspace{-2mm}
\paragraph{From ``big-bang'' updates to incremental deltas (A, B).}
A direct, monolithic prompting strategy (i.e., asking an agent to propose a complete updated architecture in one step) is brittle for the same reasons observed in requirement delta extraction: it overloads a single generation with multiple responsibilities (impact analysis, architectural consistency, interface compatibility, and syntactic correctness). Instead, we decompose the design update into a sequence of small, checkable deltas, each addressing one requirement change at a time. This mirrors industrial practice: engineers rarely accept sweeping architecture rewrites, but they can validate a small change against local constraints (interfaces, allocations, invariants) and then proceed to the next delta. Importantly, this incremental approach enables more targeted review and debugging: when a proposed update is rejected, the failure can be traced back to a specific requirement change and its local architectural implications.
The same principle carries over to verification: each delta yields a small, explicit update to monitors and tests.

\vspace{-2mm}
\paragraph{Tool-supported validation for SysML v2 model updates (C, D).}
To make incremental updates reliable, we integrate support for syntactic checks and static analyses into our toolchain for SysML~v2 artifacts. Concretely, after each proposed delta update, the updated model is recompiled to ensure valid syntax. While standard SysML~v2 tooling typically focuses on syntax checking, we additionally implement multiple domain-specific static analysis~\cite{cousot1979systematic,nielson2004principles} checkers such as \texttt{chk\_multiple\_sources\_same\_dest()} to detect cases where multiple source ports drive to the same destination port.

\begin{figure}[t]
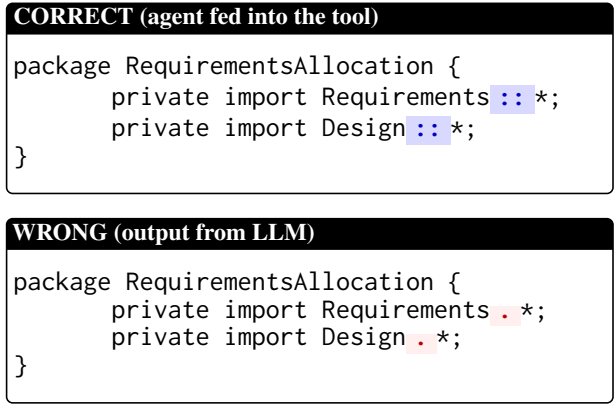

\centering

\begin{minipage}[t]{0.5\textwidth}
\begin{tcolorbox}[
  title=\textbf{CORRECT (agent fed into the tool)},
  colback=white,
  colframe=black,
  fonttitle=\bfseries\footnotesize,
  boxsep=0.5mm,
  left=0.3mm,
  right=0.1mm,
  top=0.1mm,
  bottom=0.1mm
]
\begin{lstlisting}[style=reqallocCorrect]
package RequirementsAllocation {
       private import Requirements::*;
       private import Design::*;
}
\end{lstlisting}
\end{tcolorbox}
\end{minipage}\hfill
\vspace{-2mm}
\begin{minipage}[t]{0.5\textwidth}
\begin{tcolorbox}[
  title=\textbf{WRONG (output from LLM)},
  colback=white,
  colframe=black,
  fonttitle=\bfseries\footnotesize,
  boxsep=0.5mm,
  left=0.3mm,
  right=0.1mm,
  top=0.1mm,
  bottom=0.1mm
]
\begin{lstlisting}[style=reqallocWrong]
package RequirementsAllocation {
       private import Requirements.*;
       private import Design.*;
}
\end{lstlisting}
\end{tcolorbox}
\end{minipage}
\vspace{-5mm}
\caption{Last-mile syntax drift in requirement allocation: \texttt{::} replaced by \texttt{.}, breaking compilation.}
\label{fig:reqalloc-drift}
\vspace{-1mm}
\end{figure}

In practice, however, we observed a subtle but recurring failure mode: even when the agent reports success and intermediate tool runs pass, the final step that consolidates edits into the output artifact can still introduce small formatting or structural deviations that break compilation. As an example, when the \texttt{gpt-oss:120b} is used as the LLM, in the text of requirement allocation, we observed a recurring punctuation drift where the scope-resolution token ``\texttt{::}'' is replaced by ``\texttt{.}'', as shown in Fig.~\ref{fig:reqalloc-drift}. To mitigate this, we cache the last \emph{checker-approved} model instance: whenever compilation and static analysis succeed, we persist that validated artifact. If a subsequent end-to-end generation produces an output that fails the checker, we fall back to the most recent cached, tool-validated version. This ``\emph{validated caching}'' mechanism, or checkpoint-and-rollback mechanism as termed in fault-tolerance engineering~\cite{hanmer2013patterns}, improves robustness without masking errors: it preserves the best known-good state while still surfacing the failure for diagnosis.

\vspace{-2mm}
\paragraph{Implications for test case regeneration under incremental design changes.}
Incremental architecture updates shift what must be verified: We compile each requirement/architecture delta into (i) test stimulus to be activated on the system under test, as well as (ii) formal verification constraints characterizing the correctness criteria (such as invariance conditions). These two artifacts jointly examine whether the modified system properly responds to inputs following the new requirements.

A key requirement for generating test stimuli and formal correctness criteria is \emph{the grounding in architectural interfaces}. For GenAI, the derived specification variables should be bound to concrete model elements (ports, states, or attributes), enabling direct traceability from requirement terms to the implementing design.
This yields constraints that can be translated into implementable monitors.
For example, an AEB activation requirement delta can be encoded as an invariant over explicit attribute paths (e.g., a speed range or non-faulty state), with tests generated by fixing context and perturbing only new boundary conditions.

Finally, across deltas, some changes expand the operational design domain (e.g., additional weather conditions), while others restrict it; regression shall cover combinations of accumulated refinements. As a result, the test suite scales with systematic instantiation while remaining reviewable, and each test can be traced back to the delta(s) that introduced the condition and to the exact architectural ports/states that implement it.

% \vspace{-2mm}

\section{Conclusion}
\vspace{-2mm}
In this paper, we argue that trustworthy GenAI can be achieved in a way orthogonal to foundational model development, proposing workflow-level engineering principles that remain applicable as models evolve. We showed that these principles increased robustness and traceability in automotive system engineering workflows, spanning requirements analysis, architecture updates, and verification artifacts. 
As these principles shape assurance structure rather than model capability, they can be transferred to any domain where trustworthiness is key, including safety-critical and regulated applications that require verifiable evidence and dependable outcomes.

\newpage

\section*{Ethics Statement}
The deployment of GenAI in domains such as automotive carries ethical implications regarding safety and accountability. This work specifically addresses these concerns by advocating a refined, ``checker-in-the-loop'' architecture that rejects unverified LLM outputs. Our proposed workflow is designed to augment human engineers by automating labor-intensive delta analysis and boilerplate generation, rather than replacing the expert human judgment required for final safety sign-offs. We believe this approach maintains clear lines of professional responsibility while reducing the likelihood of human fatigue-induced errors in complex requirement reviews.

\section*{Limitations}

While the proposed workflow-level design principles for trustworthy GenAI integration are generic and applicable across various system engineering domains, their implementation in production requires domain-specific adaptation. For individual applications, it remains necessary for engineers to design and validate their own sanity checks and verification rules that align with specific safety standards and local technical requirements. Our framework provides the structure for these checks, but it does not eliminate the need for expert engineering effort to define the ``ground-truth logic'' required for automated validation in different safety-critical contexts. 

\bibliography{custom}

\end{document}